\documentclass[conference]{IEEEtran}

\usepackage{makecell, lipsum}
\usepackage{enumerate}
\usepackage{tabularx}
\usepackage{booktabs}
\usepackage{threeparttable}
\usepackage{algorithmic}
\usepackage{graphicx}
\usepackage{textcomp}
\usepackage[table]{xcolor}
\usepackage[dvipsnames]{xcolor}
\usepackage{multicol}
\usepackage{multirow}
\usepackage{hyperref}
\usepackage[caption=false,font=normalsize,labelfont=sf,textfont=sf]{subfig}
\usepackage[linesnumbered,ruled]{algorithm2e}
\let\oldnl\nl
\newcommand{\nonl}{\renewcommand{\nl}{\let\nl\oldnl}}
\usepackage{pifont}
\newcommand{\cmark}{\ding{51}}%
\newcommand{\xmark}{\ding{55}}%
\usepackage{array}
\newcolumntype{C}[1]{>{\centering\arraybackslash}m{#1}}

\usepackage[all=normal, paragraphs=tight, wordspacing=tight, charwidths=normal, indent=tight, floats=tight, leading=normal, lists=tight]{savetrees}
\usepackage{float}
\AtBeginDocument{%
  \providecommand\BibTeX{{%
    Bib\TeX}}}

\usepackage{cite}
\usepackage{amsmath,amssymb,amsfonts}
\usepackage{algorithmic}
\usepackage{graphicx}
\usepackage{textcomp}
\usepackage{xcolor}
\def\BibTeX{{\rm B\kern-.05em{\sc i\kern-.025em b}\kern-.08em
    T\kern-.1667em\lower.7ex\hbox{E}\kern-.125emX}}
\begin{document}

\title{SecureRAG-RTL: A Retrieval-Augmented, Multi-Agent, Zero-Shot LLM-Driven Framework for Hardware Vulnerability Detection}


\author{
\IEEEauthorblockN{
Touseef Hasan$^{*}$,
Blessing Airehenbuwa$^{\dagger}$,
Nitin Pundir$^{\ddagger}$,
Souvika Sarkar$^{*}$,
Ujjwal Guin$^{\dagger}$
}

\IEEEauthorblockA{
$^{*}$School of Computing, Wichita State University\\
$^{\dagger}$Department of Electrical and Computer Engineering, Auburn University\\
$^{\ddagger}$IBM\\
}
}

\maketitle

\begin{abstract}
Large language models (LLMs) have shown remarkable capabilities in natural language processing tasks, yet their application in hardware security verification remains limited due to scarcity of publicly available hardware description language (HDL) datasets. This knowledge gap constrains LLM performance in detecting vulnerabilities within HDL designs. To address this challenge, we propose SecureRAG-RTL, a novel Retrieval-Augmented Generation (RAG)-based approach that significantly enhances LLM-based security verification of hardware designs. Our approach integrates domain-specific retrieval with generative reasoning, enabling models to overcome inherent limitations in hardware security expertise. We establish baseline vulnerability detection rates using prompt-only methods and then demonstrate that SecureRAG-RTL achieves substantial improvements across diverse LLM architectures, regardless of size. On average, our method increases detection accuracy by about \textit{30\%}, highlighting its effectiveness in bridging domain knowledge gaps. For evaluation, we curated and annotated a benchmark dataset of 14 HDL designs containing real-world security vulnerabilities, which we will release publicly to support future research. These findings underscore the potential of RAG-driven augmentation to enable scalable, efficient, and accurate hardware security verification workflows.
\end{abstract}

\begin{IEEEkeywords}
LLM, hardware, RAG, vulnerability detection, multi-agent systems.
\end{IEEEkeywords}

\section{Introduction}

The increased complexity of the modern semiconductor design cycle makes it difficult to ensure the security of third-party intellectual properties (IPs). Verifying IPs for security vulnerabilities requires time and subject matter expertise, which, if neglected, may result in severe security consequences~\cite{dessouky2019hardfails}. With increased design complexity, traditional simulation-based testing and formal verification \cite{drechsler2004advanced, drechsler2013formal} have become time-consuming. This underscores the need for more efficient and automated security verification methods.

Large language models (LLMs) have already demonstrated remarkable performance in the software domain for code generation~\cite{dong2024self}, understanding \cite{nam2024using}, and vulnerability detection \cite{yin2024multitask}. Recently, researchers have also started exploring LLMs in the electronic design automation (EDA) flows, such as generating hardware description language (HDL) code~\cite{thakur2023autochip, liu2023verilogeval, thakur2024verigen}, creating formal properties~\cite{hassan2024llm}, fixing syntax bugs~\cite{tsai2024rtlfixer, tarek2025socurellm}, and identifying and patching functional bugs~\cite{qayyum2024bugs, qayyum2025llm}.


\begin{table}[ht]
\centering
\fontsize{7.5}{9}\selectfont
\caption{Zero-shot prompting to detect vulnerabilities.
\vspace{-5px}
}
\label{tab:intro_table}
\setlength{\tabcolsep}{2pt}
\begin{tabular}{|c|c|c|c|c|c|c|c|c|c|c|c|c|c|c|}
\hline
\multirow{2}{*}{\textbf{Model}} & \multicolumn{14}{c|}{\makecell[c]{\textbf{Vulnerability Number}}}
\\ \cline{2-15}
 & \makecell[c]{1} & \makecell[c]{2} & \makecell[c]{3} & \makecell[c]{4} & \makecell[c]{5} & \makecell[c]{6} 
 & \makecell[c]{7} & \makecell[c]{8} & \makecell[c]{9} & \makecell[c]{10} & \makecell[c]{11} 
 & \makecell[c]{12} & \makecell[c]{13} & \makecell[c]{14} \\
\hline
Code Llama 7B It~\cite{roziere2023code} & 
\xmark & \cmark & \xmark & \xmark & \cmark & \xmark &
\cmark & \xmark & \xmark & \xmark & \cmark &
\xmark & \xmark & \xmark \\ 
StarCoder 2 7B~\cite{li2023starcoder} &
\cmark & \xmark & \xmark & \xmark & \xmark & \xmark &
\cmark & \cmark & \xmark & \xmark & \xmark &
\xmark & \cmark & \xmark \\ 
DeepSeek-Coder 6.7B It~\cite{guo2024deepseek} &
\xmark & \xmark & \xmark & \cmark & \xmark & \cmark &
\cmark & \xmark & \cmark & \xmark & \xmark &
\xmark & \xmark & \xmark \\ 
GPT-4o~\cite{hurst2024gpt} &
\cmark & \cmark & \xmark & \cmark & \xmark & \cmark &
\xmark & \xmark & \xmark & \cmark & \cmark &
\cmark & \xmark & \xmark \\ 
Gemini 2.5 Pro~\cite{comanici2025gemini} &
\cmark & \cmark & \xmark & \cmark & \xmark & \cmark &
\xmark & \cmark & \xmark & \xmark & \cmark &
\cmark & \xmark & \xmark \\ \hline
\end{tabular}
\vspace{-5px}
\end{table}

Commercial LLMs are trained on huge corpora of web data, and their training data heavily suffers from an under-representation of HDLs. Even LLMs trained for code-specific tasks contain GitHub repositories mostly written in C, Python, Java, etc., because of their widespread availability online \cite{xu2022systematic}. This could potentially hamper the hardware understanding capabilities of the LLMs. Our initial experiments reflected the same in Table~\ref{tab:intro_table}, where code-based LLMs like Code Llama~\cite{roziere2023code}, StarCoder~\cite{li2023starcoder}, DeepSeek-Coder~\cite{guo2024deepseek} could only detect 4 out of the 14 vulnerabilities in the HDL testset, and frontier LLMs like GPT-4o~\cite{hurst2024gpt}, Gemini 2.5 Pro~\cite{comanici2025gemini} could only detect 7. This represents how far LLMs currently are from robust vulnerability detection in HDLs. Although researchers have recently demonstrated success in hardware vulnerability detection by fine-tuning pre-trained LLMs~\cite{fu2023llm4sechw, yao2025location, tarek2025bugwhisperer, yao2024hdldebugger}, the associated computational costs raise questions regarding their wide adoption. 

In this paper, we introduce \textbf{SecureRAG-RTL}, a novel retrieval-augmented, multi-agent, zero-shot hardware vulnerability detection flow to achieve significantly improved detection performance across diverse LLM families. The proposed framework detects all 14 benchmarked RTL vulnerabilities in the test set, while demonstrating increased accuracy for small and medium-sized LLMs. Our key contributions are summarized as follows:

\vspace{-5px}

\begin{itemize}
    \item We propose \textbf{SecureRAG-RTL}, a multi-agentic zero-shot prompting flow designed to enhance vulnerability detection in HDL without requiring LLM fine-tuning.

    \item Our proposed novel RAG pipeline integrates structured CWE knowledge for efficient and semantically rich vulnerability search. Our approach leverages RTL signature extraction, multi-field semantic search, HDL annotation, and LLM-assisted RTL summarization. Together, these significantly improve the precision and contextual relevance of CWE-based vulnerability detection in hardware designs.
    
    \item We evaluate 18 open-source and proprietary LLMs for vulnerability detection in HDL code using \textbf{SecureRAG-RTL}, showing consistent improvements (up to \textit{42\%}). The proposed security verification pipeline is specifically designed to operate efficiently with lightweight models, focusing on resource constraints and emphasizing on localized security verification, accommodating the proprietary nature of HDL in industrial settings.
\end{itemize}

The rest of the paper is organized as follows. Section \ref{sec:prior-work} includes the background study on hardware vulnerabilities, LLMs for vulnerability detection, and RAG. We present the SecureRAG-RTL framework in Section~\ref{section:approach}. Section \ref{section: experiments} covers the details of our experimental design. Section \ref{section: discussion} contains the results and summarizes our findings. Section \ref{section: conclusion} concludes the paper.

\vspace{-5px}

\section{Prior Work}
\label{sec:prior-work}

This section explores the application of LLMs in hardware vulnerability detection and highlights the potential of RAG. 

\vspace{-5px}
\subsection{LLMs for HDL Vulnerability Detection}
\label{subsec:vul-detection}

Researchers have demonstrated the automatic generation of hardware code using LLMs by fine-tuning pre-existing models on HDL datasets~\cite{thakur2024verigen, fu2024hardware} and shown comparable results to frontier LLMs. However, using LLMs to detect vulnerabilities in HDLs hasn't been a straightforward or successful venture. There is no one-size-fits-all solution, and so prior works have explored various approaches. For instance, RTLFixer~\cite{tsai2024rtlfixer} utilizes LLMs to autonomously plan intermediate steps for iterative debugging, thereby fixing syntax errors in HDL. VeriAssist~\cite{huang2024towards} enables LLMs to self-correct their generated code by adopting an automatic prompting system, which involves multi-turn interactions and chain-of-thought reasoning. Self-HWDebug~\cite{akyash2024self} leverages LLMs to automatically create required debugging instructions, while another study proposes an ensemble of LLMs to generate and evaluate repairs~\cite{ahmad2024hardware}. MARVEL~\cite{collini2025marvel} and MEIC~\cite{xu2024meic} make use of multiple LLM agents for debugging RTL security bugs. Another approach proven to be effective but expensive is fine-tuning LLMs specifically for vulnerability detection tasks in HDL code, as seen in LLM4SecHW~\cite{fu2023llm4sechw}, LiK~\cite{yao2025location}, BugWhisperer~\cite{tarek2025bugwhisperer}, and HDLdebugger~\cite{yao2024hdldebugger}. 
Despite these attempts, many of these approaches rely on fine-tuning, which is too costly and impractical for low-resource environments.

\vspace{-5px}
\subsection{Hardware Vulnerability Database}
\label{subsec:hard-cvd}

Hardware vulnerabilities are weaknesses or flaws within the physical, logical, or architectural design of computing systems that can be exploited to undermine a system's integrity, confidentiality, or availability. Unlike software bugs, hardware vulnerabilities are often more difficult to detect and mitigate, as they become embedded during the design and fabrication process~\cite{dessouky2019hardfails}. Recently, a list of hardware-specific vulnerabilities has been added to the Common Weakness Enumeration (CWE) list hosted by the MITRE corporation~\cite{mitrecorp}. The presence of a weakness indicates that there are “flaws, faults, bugs, or other errors in software or hardware implementation, code, design, or architecture that if left unaddressed could result in systems, networks, or hardware being vulnerable to attack”. This database serves as a common reference for RTL designers to navigate hardware weaknesses.

\begin{figure*}    \centerline{\includegraphics[width=1\linewidth]{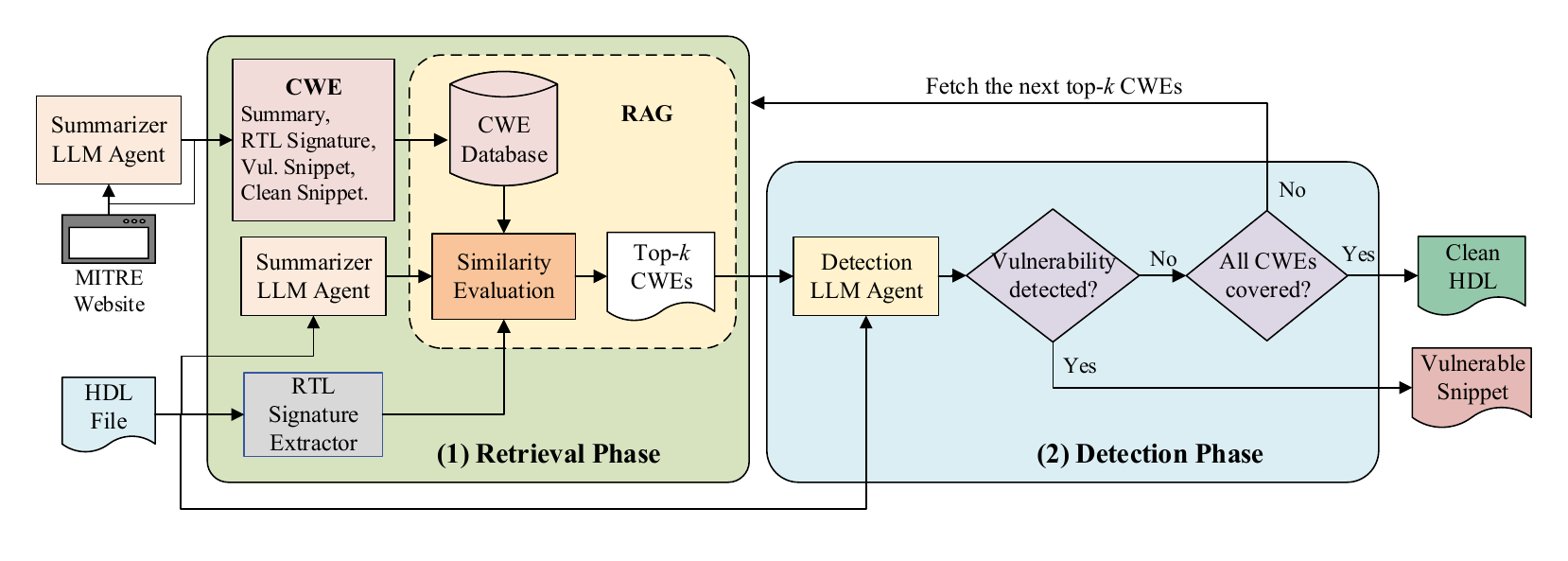}} \vspace{-20px}
    \caption{The proposed SecureRAG-RTL pipeline for hardware vulnerability detection.} \vspace{-10px}
    \label{fig:overallworkflow}
\end{figure*}

\vspace{-5px}

\subsection{Retrieval-Augmented Generation}
\label{sub-section: RAG}

LLMs are trained on exhaustive web corpora but often hallucinate due to a lack of domain-specific context. To overcome this limitation, retrieval-augmented generation (RAG)~\cite{lewis2020retrieval} allows the LLM to access and retrieve relevant documents from an external knowledge base. This improves the accuracy and factual consistency of the model in generating the desired output. RAG is beneficial in our context since most state-of-the-art code-based LLMs are pre-trained on GitHub repositories dominated by C, Python, and Java~\cite{xu2022systematic} and thus, HDLs such as Verilog are severely underrepresented in these training sets. This augmentation is also particularly useful for smaller open-source models, whose base performance is insufficient for practical deployment in hardware security pipelines. Appending hardware-specific knowledge in an external knowledge base can provide the LLMs with more context regarding RTL designs and potentially improve vulnerability detection rates. In the context of vulnerability detection in RTL code via LLMs, several works~\cite{qayyum2024bugs, qayyum2025llm} have utilized the RAG mechanism to classify and patch different functional bugs iteratively. RTLFixer~\cite{tsai2024rtlfixer} stores human expertise in a retrieval database to provide guidance and explanations whenever LLMs face difficulty in patching bugs. Another study makes use of a database containing content from HDL books and achieves promising results in root cause analysis of hardware design failures using frontier LLMs~\cite{qiu2025towards}. Despite these advances, none of these methods systematically use CWE metadata to retrieve relevant weaknesses and guide LLMs to accurately identify the presence of vulnerabilities in hardware designs.

\vspace{-3px}
\section{The SecureRAG-RTL Framework}
\label{section:approach}

This section presents our proposed SecureRAG-RTL framework for detecting vulnerabilities in RTL code based on a multi-agent, zero-shot, retrieval-augmented approach. As illustrated in Figure~\ref{fig:overallworkflow}, the framework operates in two phases, namely: (i) \textbf{Retrieval Phase}, where CWEs semantically similar to HDL summaries are retrieved from the CWE knowledge database using cosine similarity, and (ii) \textbf{Detection Phase}, where the RTL designs are iteratively evaluated using the retrieved CWE context to identify security vulnerabilities and extract vulnerable snippets from the designs.

\vspace{-5px}

\subsection{Retrieval Phase}
\label{subsection: retrieval}

This phase is the first phase of the SecureRAG-RTL framework. Here, the system identifies CWEs relevant to the input RTL design via semantic similarity. The framework computes the cosine similarity between the embedding vectors of the RTL, its summary, and signatures to those of CWEs stored in the knowledge database, ensuring that only the most contextually relevant vulnerabilities are retrieved for downstream analysis. The process is decomposed into three major sub-steps, as detailed below and algorithmically represented in Algorithm~\ref{alg:cwe_ingestion}. The algorithm is described as follows:


\begin{algorithm}
\caption{Top CWEs Retrieval}
\label{alg:cwe_ingestion}
\begin{algorithmic}[1]
\REQUIRE \texttt{SoC}: HDL, \texttt{$CWE = \{cwe_1,cwe_2,...,cwe_n\}$}: List of CWEs, $A_{S}$: Summarizer Agent, $KW_E$: Keyword Extractor, $\mathcal{DB}$: CWE Vector Database, $P_{VS}$: Vulnerable and Secure Snippet Prompt, $P_{SK}$: Summarizer and Keyword Prompt, $P_{S}$: Summarizer Prompt, $Emb$: Text Embedding, $R_{topK}$: Retrieve Top K.
\ENSURE \texttt{$TpKCWE = \{cwe_1, cwe_2, cwe_3, cwe_4, cwe_5, ... cwe_k\}$}
\small{\color{blue}/* CWE KDB Construction */}
\STATE Initialize $\mathcal{DB} \gets \emptyset$, $metadata \gets \emptyset$
\FORALL{$cwe_i \in CWE$}
    \STATE $ID_i,Title_i,Desc_i,ExtDesc_i \gets cwe_i$
    \STATE $S_i,Kw_i \gets A_{S}(P_{SK}, Desc_i, ExtDesc_i)$ \hfill \small{\color{blue}/* CWE summary  and Keywords*/}
    \STATE $VS_{i},SS_{i} \gets A_{S}(P_{VS}, Desc_i, ExtDesc_i)$  \hfill \small{\color{blue}/* Vulnerable and secure snippets*/}
    \STATE $metadata \gets metadata \cup \{ID_i, Title_i, S_i, Kw_i,VS_i,SS_i\}$
\ENDFOR
\STATE $embed \gets Emb(metadata)$
\STATE $\mathcal{DB} \gets \{metadata, embed\}$ \hfill \small{\color{blue}/* Add to DB */} \;
\small{\color{blue}/*HDL Keywords and Summary */ } 
\STATE $S \gets A_{S}(P_{S}, SoC)$
\STATE $KW \gets KW_{E}(SoC)$
\small{\color{blue}/* Top-K CWE Retrieval */} 
\STATE $Q \gets \{\alpha\times Emb(S)+\beta\times Emb(KW)\}$ \hfill \small{\color{blue} /* Query */}
\STATE \texttt{TpKCWE} $\gets R_{topK}(\mathcal{DB}, Q)$
\RETURN \texttt{TpKCWE}
\end{algorithmic}
\vspace{-5px}
\end{algorithm}

\noindent Line 1 initializes the CWE knowledge database $\mathcal{DB}$, which stores all hardware-related CWE entries. From the list of CWEs obtained from MITRE, a set of enhanced metadata is constructed (Line 2--7). Each metadata record includes a CWE summary, a set of keywords (Hardware Signature), and paired vulnerable and secure HDL examples. Multi-fielded storage of CWEs is essential for effective retrieval in RAG and semantic search systems. Ingesting long, unsegmented documents can dilute the significance of critical terms, making it harder to retrieve relevant CWEs. By segmenting each CWE into smaller and focused fields, semantic comparison becomes more precise and less noisy. Additionally, field-level granularity enables the retrieval system to better interpret the role and context of information. For instance, a hardware signature containing terms like [JTAG, Key] immediately signals that the corresponding hardware involves a JTAG structure and the use of a confidential key. A context that would require a lot more tokens and would be difficult to extract from a monolithic document.

The CWEs metadata are generated by the Summarizer LLM Agent based on the raw CWE description and extended description from MITRE, while preserving the raw CWE ID and title. Line 8 then computes sentence-level embeddings for each metadata entry, enabling efficient retrieval based on embeddings. Once the embeddings are computed, both the enhanced metadata and their corresponding embeddings are ingested into $\mathcal{DB}$ for downstream querying and analysis (Line 9).

For a given HDL input, a Summarizer Agent generates a technical summary $S$ for the HDL in Line 10. Line 11 extracts hardware signature $KW$ (e.g., DRAM, Key, JTAG, UART, and Auth) from the HDL using a Python-based parser $KW_{E}$. Once the $S$ and $KW$ are obtained, a weighted sum of their embeddings is computed to get the query embedding $Q$, which represents the semantic profile of the HDL potential vulnerabilities in Line 12. To determine the appropriate values for the weighting coefficients, we performed experiments that varied the contributions of the HDL summary and hardware signature embeddings. The combination of $\alpha = 0.7$ and $\beta =0.3$ yielded the highest retrieval performance. 
Finally, in Line 13, given the CWE knowledge database $\mathcal{DB}$ and query $Q$, the function $R_{topK}$ performs similarity search based on cosine similarity over the vector database and returns the Top-$k$ CWEs related to the query in \texttt{TpKCWE}.

\subsubsection{RTL Summary and Signature Extraction}\label{subsubsec:rtl_analysis}
The input RTL design undergoes semantic transformation using an LLM agent, referred to as~\textit{Summarizer Agent}. Given the RTL design, the Summarizer Agent generates a technical summary that encapsulates key aspects, including design intent, FSM structure, and data/control dependencies from the design. In parallel, a Python-based parser extracts relevant keywords directly from the source to constitute the hardware signature. The output response of the Agent and the extracted signature are embedded into a vector representation, serving as a semantic query for CWE retrieval from the database.

\subsubsection{CWE Knowledge Database}
The CWE knowledge database serves as a searchable repository of hardware-related CWEs. To build this database, we first extract relevant CWE entries from the MITRE website using web scraping. For each CWE, key metadata fields, including the CWE-ID, title, description, extended description, mitigation, and modes of introduction, are collected. From this raw metadata, we employ the Summarizer Agent in processing the CWE description, extended description, mitigation, and modes of introduction to construct an enriched metadata set that includes CWE-ID, Title, LLM-generated keywords, a pair of LLM-generated vulnerable and secure snippets that demonstrate the vulnerability, and an LLM-generated summary that describes the CWE. This step is necessary to ensure consistency in how the CWEs are written, since the original descriptions provided by MITRE are human-written and may differ in style or level of detail, and also because the RTL summary described in Section~\ref{subsubsec:rtl_analysis} is LLM-generated. Aligning both sides using the same generation pipeline enhances semantic comparability during the retrieval phase~\cite{karev2025conscompf}. 

\subsubsection{Hybrid Retrieval and Evaluation}
Once the natural language summaries and keywords are extracted from the input RTL, we retrieve the most relevant hardware CWEs from a semantically enriched CWE database by a dense embedding-based similarity search. The summary and hardware signature embeddings generated from the input RTL are compared to the embeddings of CWE metadata (i.e., LLM-generated summary, vulnerable snippet, and keywords) using the cosine similarity score.
Cosine similarity measures the semantic overlap of by encoding tokens into dense vector embeddings. Even if the phrasing differs, cosine similarity evaluates if the semantic meaning is preserved. This metric evaluates the semantic overlap between the CWE metadata and the HDL files.

The objective of retrieving the most relevant CWEs is to form a context that will assist the agent in the detection phase in classifying the given RTL as either vulnerable or secure and to identify the CWEs present in the code. The LLM agent will essentially get a narrowed-down pool of CWEs and reason why the given RTL is vulnerable to the predicted CWE(s). We note that our framework is designed to retrieve the top-10 most relevant CWEs in terms of semantic similarity with the input RTL code. However, this framework can be adapted to retrieve the top-$k$ most relevant CWEs, depending on how many similar CWEs are helpful for the agent(s) in the detection phase to predict the accurate ones.

\begin{figure}[t]
\centerline{\includegraphics[width=1\linewidth]{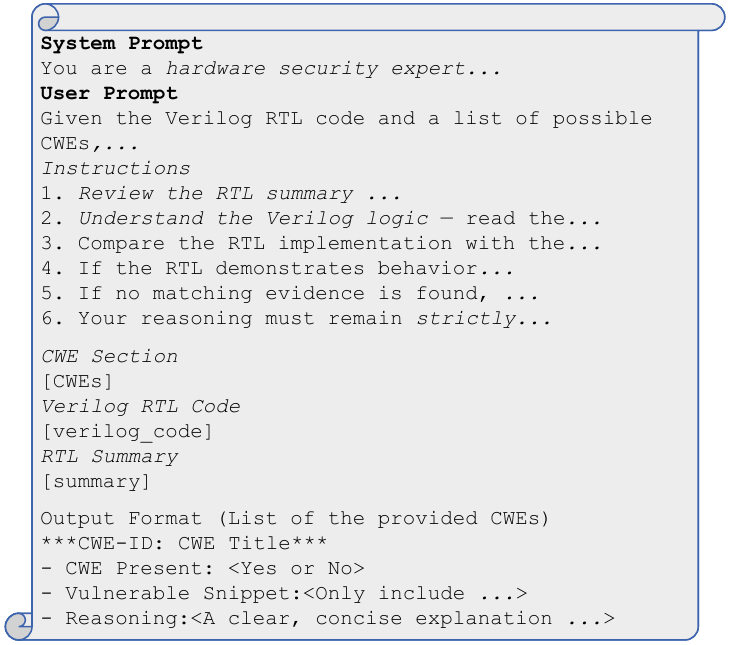}\vspace{-10px}}
    \caption{Detection Agent Prompt.}
    \label{fig:prompt}
    \vspace{-10px}
\end{figure}

\vspace{-5px}
\subsection{Detection Phase}
\label{subsection: detection}
\vspace{-3px}
In this phase, we employ \textit{Detection Agent}, which is the core analytical component responsible for identifying security vulnerabilities within the RTL designs. As illustrated in Figure~\ref{fig:prompt}, the agent operates using a carefully structured prompt that guides it to reason about hardware vulnerabilities in a manner similar to that of a human hardware security expert.
The role of a hardware security expert is assigned to the agent and enriched with the RTL summary and top-$k$ retrieved CWE details from the retrieval phase. Due to context window limitations, smaller models such as Code Llama process the retrieved CWEs iteratively, analyzing one CWE at a time. In contrast, larger-capacity models can accommodate the full top-$k$ CWE set within a single prompt, enabling a more comprehensive and holistic evaluation. The complete Detection prompt is  at~\cite{iyobosa25}.

\begin{figure}
    \centerline{\includegraphics[width=1\linewidth]{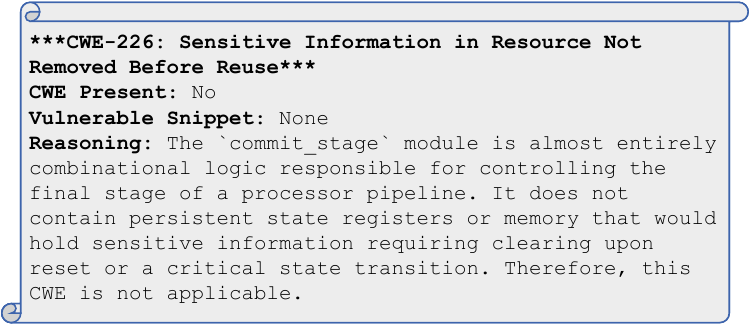}} \vspace{-10px}
    \caption{Detection agent responses when CWE is not found.}
    \label{fig:response2}
\end{figure}

\begin{figure}
\centering
\includegraphics[width=1\linewidth]{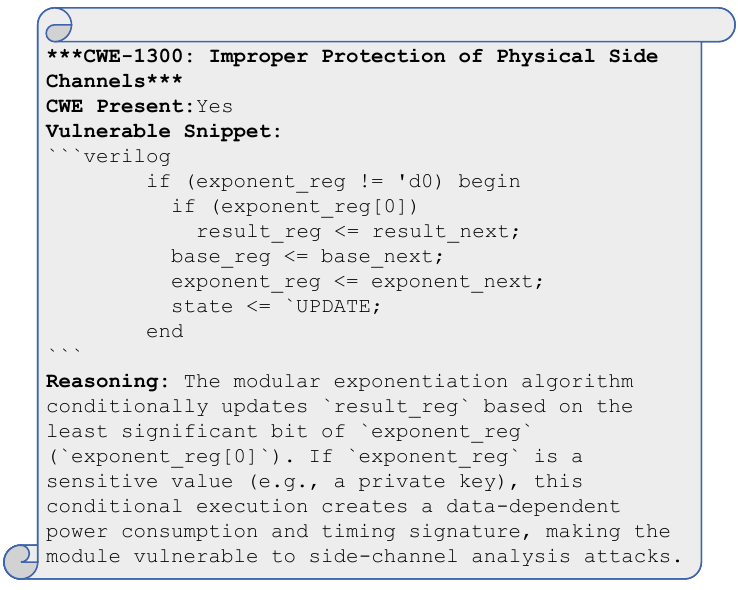} \vspace{-20px}
    \caption{Detection agent responses when a CWE is found.}
    \label{fig:response}
\end{figure}

Adding this context enables the agent to systematically examine the RTL, interpret its design intent, and implementation to find indicators associated with each of the retrieved CWEs. Where the given CWE does not exist in the RTL, the agent outputs a response as given in Figure~\ref{fig:response2}. For each CWE identified to exist in the RTL by the agent, the Detection Agent extracts that section of the code with the potential vulnerability as demonstrated by the agent's response in Figure~\ref{fig:response}. The correctness of this snippet is evaluated using the ROUGE-L metric, discussed in Section~\ref{subsection: eval}.

\vspace{-5px}
\section{Experimental Setup}
\label{section: experiments}

\vspace{-3px}
\subsection{Dataset Construction}

Our dataset is derived from the HDL modules provided in Hack@DA-C'21~\cite{hackdac}. We selected modules for which the vulnerable code regions and corresponding fixes were available. To expand the dataset and increase vulnerability diversity, we additionally injected security flaws into clean Hack@DAC’21 modules, resulting in a dataset spanning 14 distinct vulnerability types, with each entry in the dataset consisting of: (1) the buggy HDL, (2) the gold-standard vulnerable snippet, and (3) the corrected version of the design.

\vspace{-3px}
\subsection{Models Evaluated}

To evaluate the effectiveness of our approach, we conducted experiments using a total of 18 LLMs of varying nature and parameter sizes. We experimented with open-soured LLMs, such as code-based models like Code Llama \cite{roziere2023code}, Qwen Coder \cite{hui2024qwen2}, StarCoder \cite{li2023starcoder}, DeepSeek-Coder \cite{guo2024deepseek}, and reasoning models like DeepSeek-R1 \cite{guo2025deepseek}, Falcon \cite{almazrouei2023falcon}, Granite \cite{granite2024granite}, Gemma \cite{team2025gemma}, Phi \cite{abdin2024phi3technicalreporthighly}, Mistral \cite{jiang2023mistral7b}, Qwen \cite{yang2025qwen3}, and Llama \cite{grattafiori2024llama}. Finally, we tested different frontier state-of-the-art (SoTA) LLMs like ChatGPT \cite{hurst2024gpt} and Gemini \cite{comanici2025gemini}.

We accessed the proprietary models using their respective APIs except for GPT-5, for which we used OpenAI’s web interface, as an API for this model was not available at the time of experimentation. All the open-sourced models were accessed via HuggingFace pipelines~\cite{huggingface}. Our RAG pipeline used the \textit{ChromaDB} vector database and the \textit{all-mpnet-base-v2} sentence embedding model. In creating the CWE knowledge database, a total of 50 hardware-related CWEs were web-scraped from the MITRE website, and a pair of vulnerable and secure snippets was generated for each CWE using GPT-5.

\vspace{-5px}
\subsection{Evaluation Metric}
\label{subsection: eval}

To assess the effectiveness of the framework in identifying the vulnerable code snippets from the HDL modules, we employ the Rouge-L score. Rouge-L measures the longest common subsequence (LCS) between the true and predicted labels as denoted in Equation \ref{eq:eq1}. 

\vspace{-5px}
\begin{equation}
F_{\text{LCS}} = \frac{(1 + \beta^2) \cdot P_{\text{LCS}} \cdot R_{\text{LCS}}}{R_{\text{LCS}} + \beta^2 \cdot P_{\text{LCS}}} \label{eq:eq1}
\end{equation}

where, $P_{\text{LCS}}$ and $R_{\text{LCS}}$ denote the precision and recall respectively. We set $\beta = 1$ to assign equal importance to both precision and recall, as our objective is to ensure that the LLM provides the HDL snippets that are not only accurate but also complete in representing the intended vulnerabilities. In contrast, $\beta > 1$ is commonly adopted in summarization tasks, where recall is prioritized to maximize coverage of reference content.




\section{Result and Discussion}
\label{section: discussion}

The evaluation of the SecureRAG-RTL framework across various LLM sizes reveals that small models benefit the most from our approach, highlighting its potential for practical hardware-security deployment in resource-constrained environments.

\begin{table}[t]
\centering
\fontsize{7.5}{8.5}\selectfont
\caption{Retrieval phase evaluation of different models.}
\vspace{-5px}
\begin{tabular}{|p{1.1cm}|p{4.2cm}|C{0.4cm}|C{0.4cm}|C{0.4cm}|}
\hline
\textbf{Category} & \textbf{Models Experimented} & \textbf{T1} & \textbf{T5} & \textbf{T10} \\
\hline
\multirow{6}{*}{\makecell{Small \\ Models}} 
& DeepSeek-R1-Distill-Qwen 1.5B & 0 & 2 & 3 \\
& Qwen 2.5 Coder 3B Instruct      & 1 & 3 & 4 \\
& Falcon 3 3B Instruct           & 2 & 2 & 6 \\
& Granite 4.0 Micro 3B           & 2 & 4 & 6 \\
& Phi 3 Mini 3.8B Instruct       & 1 & 2 & 4 \\
& Gemma 3 4B Instruct            & 2 & 4 & 7 \\
\hline
\multirow{6}{*}{\makecell{Medium \\ Models}}
& Code Llama 7B Instruct         & 3 & 5 & 6 \\
& StarCoder 2 7B                 & 4 & 6 & 7 \\
& DeepSeek-Coder 7B Instruct     & 4 & 6 & 7 \\
& Mistral v0.3 7B                & 2 & 5 & 5 \\
& Qwen 3 8B                      & 3 & 6 & 8 \\
& LLaMA 3.2 11B Vision Instruct   & 4 & 7 & 8 \\
\hline
\multirow{6}{*}{\makecell{SoTA \\ Models}} 
& Gemini 2.0 Flash               & 6 & 11 & 12 \\ 
& Gemini 2.5 Flash Lite          & 7 & 10 & 13 \\
& Gemini 2.5 Flash               & 6 & 10 & 13 \\
& Gemini 2.5 Pro                 & 5 & 13 & 14 \\ 
& GPT-4o                         & 6 & 11 & 12  \\
& GPT-5                          & 6 & 12 & 12 \\
\hline
\end{tabular}
\vspace{-5px}
\label{tab:retrievedCWEs}
\end{table}

To further understand the performance of LLMs across different size categories, Table \ref{tab:retrievedCWEs} compares top 1 (T1), top 5 (T5), and top 10 (T10) retrieval accuracy during the CWE retrieval phase across small, medium, and state-of-the-art (SoTA) LLMs. Although many LLMs struggled to identify the correct CWE entries from the knowledge base, the Gemini and GPT models retrieved the correct CWE in the T5 candidates for 10-13 out of 14 test cases. Notably, Gemini 2.5 Pro retrieved the correct CWE in all 14 test cases when the T10 candidates were considered. In contrast, most other LLMs, particularly the small and medium-sized models, barely managed to retrieve the correct CWE in just half of the 14 test cases in their T10 retrieval. This highlights a crucial limitation: small and medium models lack fundamental hardware knowledge to generate strong and relevant RTL summaries and signatures for effective semantic searching. SecureRAG-RTL circumvents this limitation by delegating the RTL summary and signature generation task for metadata to a SoTA model (e.g., Gemini 2.5 Pro). This is a one-time hit taken to effectively guide small models in vulnerability detection.

\begin{table*}[htbp]
\centering
\fontsize{7.5}{8.5}\selectfont
\caption{A comparative visualization of the ROUGE-L scores (in \%) before and after incorporating SecureRAG-RTL.} 
\vspace{-5px}
\renewcommand{\arraystretch}{1}
\setlength{\tabcolsep}{5pt}
\begin{tabularx}{\textwidth}{l *{14}{c}}
\hline
\multirow{2}{*}{\textbf{Model}} & \multicolumn{14}{c}{\makecell[c]{\textbf{Vulnerability Number}}}
\\ \cline{2-15}
 & \makecell[c]{1} & \makecell[c]{2} & \makecell[c]{3} & \makecell[c]{4} & \makecell[c]{5} & \makecell[c]{6} 
 & \makecell[c]{7} & \makecell[c]{8} & \makecell[c]{9} & \makecell[c]{10} & \makecell[c]{11} 
 & \makecell[c]{12} & \makecell[c]{13} & \makecell[c]{14} \\
\hline
DeepSeek-R1-D-Q 1.5B & 0/0 & 11/24 & 3/5 & 59/71 & 4/5 & 14/14 & 0/0 & 16/18 & 0/3 & 0/0 & 9/11 & 1/10 & 7/15 & 0/0 \\
Qwen 2.5 Coder 3B It & 4/54 & 0/2 & 54/60 & 9/38 & 59/68 & 0/9 & 62/65 & 9/58 & 11/14 & 2/3 & 15/68 & 0/0 & 7/9 & 0/1 \\
Falcon 3 3B It & 54/55 & 0/61 & 0/0 & 0/1 & 3/10 & 67/73 & 9/31 & 14/15 & 16/31 & 9/9 & 5/10 & 0/1 & 0/51 & 0/0 \\
Granite 4.0 Micro 4B & 60/67 & 48/53 & 0/2 & 7/10 & 5/32 & 13/13 & 12/12 & 0/45 & 2/3 & 9/59 & 1/1 & 7/62 & 12/13 & 0/4 \\
Phi 3 Mini 3.8B It & 1/2 & 0/0 & 13/28 & 69/81 & 0/8 & 7/10 & 10/79 & 11/12 & 0/0 & 8/12 & 4/5 & 1/1 & 9/13 & 0/5 \\
Gemma 3 4B It & 0/2 & 6/6 & 92/93 & 64/100 & 2/30 & 0/2 & 9/52 & 12/40 & 46/59 & 0/80 & 11/47 & 2/39 & 3/10 & 8/8 \\
Code Llama 7B It & 2/28 & 56/62 & 10/12 & 0/36 & 34/40 & 8/9 & 48/94 & 9/52 & 7/9 & 0/2 & 35/68 & 4/9 & 12/12 & 0/4 \\
StarCoder 2 7B & 59/67 & 0/34 & 13/13 & 0/0 & 9/12 & 4/42 & 38/56 & 49/69 & 0/0 & 2/4 & 10/10 & 0/10 & 61/88 & 0/0 \\
DeepSeek-Coder 7B It & 3/8 & 7/12 & 10/60 & 56/75 & 0/0 & 40/82 & 0/2 & 72/100 & 10/10 & 6/7 & 0/32 & 12/12 & 2/10 & 0/27 \\
Mistral v0.3 7B & 8/76 & 1/12 & 0/0 & 45/100 & 0/38 & 12/16 & 50/81 & 39/42 & 1/2 & 7/56 & 18/18 & 0/0 & 11/17 & 0/12 \\
Qwen 3 8B & 19/38 & 26/64 & 0/2 & 45/98 & 2/56 & 13/43 & 39/78 & 45/100 & 0/0 & 6/9 & 10/33 & 20/39 & 76/100 & 0/12 \\
LLaMA 3.2 11B V-It & 26/33 & 10/12 & 0/8 & 41/89 & 65/77 & 0/48 & 71/71 & 15/28 & 0/0 & 2/34 & 38/90 & 1/64 & 0/47 & 0/11 \\
Gemini 2.0 Flash & 35/49 & 24/100 & 0/15 & 18/18 & 0/24 & 0/62 & 0/35 & 30/30 & 5/100 & 0/29 & 100/100 & 56/56 & 0/0 & 0/0 \\
Gemini 2.5 Flash Lite & 0/39 & 30/33 & 0/3 & 0/9 & 0/83 & 0/100 & 0/40 & 30/89 & 0/40 & 0/58 & 100/100 & 56/56 & 0/0 & 0/0 \\
Gemini 2.5 Flash & 0/56 & 0/24 & 0/100 & 18/18 & 100/100 & 0/100 & 27/34 & 20/100 & 2/12 & 34/58 & 53/100 & 33/41 & 0/100 & 0/72 \\
Gemini 2.5 Pro & 20/49 & 58/58 & 100/100 & 18/18 & 4/4 & 70/70 & 27/34 & 20/100 & 2/20 & 58/58 & 44/92 & 33/35 & 1/74 & 0/76 \\
GPT-4o & 75/79 & 69/70 & 3/3 & 55/60 & 11/12 & 89/100 & 0/88 & 16/59 & 6/61 & 82/82 & 100/100 & 100/100 & 2/84 & 0/91 \\
GPT-5 & 74/81 & 48/48 & 79/84 & 69/88 & 70/82 & 0/21 & 30/36 & 39/39 & 2/39 & 11/11 & 7/30 & 90/100 & 2/35 & 0/0 \\
\hline
\end{tabularx}
\label{tab: rouge}
\vspace{-15px}
\end{table*}

The effectiveness of this approach becomes apparent in the detection accuracy results illustrated in Table \ref{tab:differentmodelresults}. Before applying our SecureRAG-RTL approach, most small and medium-scale LLMs performed poorly on detecting vulnerabilities in HDL files. Detection accuracy for small models ($\leq$4B) like DeepSeek-R1-Distill-Qwen 1.5B, Phi 3 Mini 3.8B, and Gemma 3 4B hovered around 7\% to 21\%. These scores suggest that without domain-specific context, small models struggle in the vulnerability detection task, likely due to a lack of training on RTL patterns and hardware-specific semantics. After applying SecureRAG-RTL, their detection accuracy improved drastically. For example, Gemma 3 4B jumped from 21.42\% to 64.29\%, tripling its detection accuracy. Similarly, the rest of the small models also achieved a detection accuracy of thrice on average, compared to pre-RAG approaches. This shows that even if a one-time cost is incurred in using a large model during ingestion, our framework enables low-resource LLMs to improve their vulnerability detection capabilities.

The performance of medium-sized models ($>$4B, $\leq$11B), such as DeepSeek-Coder 7B Instruct and Code Llama 7B Instruct, followed a similar trend. Prior to RAG, these models LLMs achieved moderate accuracies between 28\% and 35\%. But when paired with SecureRAG-RTL, their detection improved significantly. For instance, Mistral v0.3 7B and Qwen 3 8B both almost doubled their detection accuracy to 42.8\% and 71.4\%, respectively. These improvements suggest that SecureRAG-RTL benefits not just small models but also mid-tier models by providing them with semantic scaffolding.

\begin{table}[t]
\small
\centering
\fontsize{7.5}{8.5}\selectfont 
\caption{Comparison of detection accuracy before and after implementing SecureRAG-RTL.} \vspace{-5px}
\begin{tabular}{|l|c|c|c|}
\hline
\textbf{Models Experimented} 
& \makecell{\textbf{Accuracy}\\\textbf{(Before)}} 
& \makecell{\textbf{Accuracy}\\\textbf{(After)}} 
& \makecell{\textbf{Increase}} \\
\hline
DeepSeek-R1-D-Qwen 1.5B & 7.14\% & 21.43\% & 14.29\% \\
Qwen 2.5 Coder 3B It & 21.43\% & 50.00\% & 28.57\% \\
Falcon 3 3B It & 14.29\% & 42.86\% & 28.57\% \\
Granite 4.0 Micro 3B & 14.29\% & 42.86\% & 28.57\% \\
Phi 3 Mini 3.8B It & 7.14\% & 21.43\% & 14.29\% \\
Gemma 3 4B It & 21.43\% & 64.29\% & 42.86\% \\
\hline
\rowcolor{gray!10}
\textbf{Avg. for small LLMs} & \textbf{14.29\%} & \textbf{40.48\%} & \textbf{26.19\%} \\
\hline
Code Llama 7B It & 28.57\% & 50.00\% & 21.43\% \\
StarCoder 2 7B & 28.57\% & 42.86\% & 14.29\% \\
DeepSeek-Coder 7B It & 28.57\% & 42.86\% & 14.29\% \\
Mistral v0.3 7B & 21.43\% & 42.86\% & 21.43\% \\
Qwen 3 8B & 35.71\% & 71.43\% & 35.71\% \\
LLaMA 3.2 11B V-It & 35.71\% & 64.29\% & 28.57\% \\
\hline
\rowcolor{gray!10}
\textbf{Avg. for medium LLMs} & \textbf{29.76\%} & \textbf{52.38\%} & \textbf{22.62\%} \\
\hline
Gemini 2.0 Flash & 50.00\% & 85.71\% & 35.71\% \\ 
Gemini 2.5 Flash Lite & 35.71\% & 85.71\% & 50.00\% \\
Gemini 2.5 Flash & 57.14\% & 100.00\% & 42.86\% \\
Gemini 2.5 Pro & 50.00\% & 100.00\% & 50.00\% \\ 
GPT-4o & 50.00\% & 85.71\% & 35.71\% \\
GPT-5 & 57.14\% & 85.71\% & 28.57\% \\
\hline
\rowcolor{gray!10}
\textbf{Avg. for the SoTA LLMs} & \textbf{50.00\%} & \textbf{90.48\%} & \textbf{40.48\%} \\
\hline
\end{tabular}
\label{tab:differentmodelresults}
\vspace{-8px}
\end{table}

Even the frontier LLMs, i.e., the GPT and Gemini models, showed substantial benefits via SecureRAG-RTL. While these LLMs already achieved baseline detection accuracies without RAG, our framework pushed their accuracies to over 85\%, with Gemini 2.5 Flash and Pro notably reaching 100\%, detecting all the vulnerabilities successfully. This indicates RAG remains beneficial even for state-of-the-art models trained on exhaustive code and language corpora.

Beyond detection, we also analyzed the quality of the model outputs using the ROUGE-L score, which measures how exactly the models could extract the vulnerable snippets. After integrating our framework, ROUGE-L scores increased significantly, as evident in Table \ref{tab: rouge}. In smaller models, the increase in scores highlighted that our framework enabled them to not only detect more vulnerabilities, but also extract vulnerable snippets from HDL code more precisely than before. In larger models, the ROUGE-L scores approached 90\% for many test cases after implementing our framework.

One of the most compelling findings is that the magnitude of improvement due to SecureRAG-RTL was inversely proportional to model size, i.e., smaller LLMs benefit more. This suggests that SecureRAG-RTL acts as a knowledge amplifier, particularly for models that are not pretrained on specialized code corpora. Instead of relying on resource-intensive models like GPT-5 or Gemini 2.5 Pro for detection, 3-4B models can be used with negligible loss in accuracy, especially when combined with retrieval phases using larger models. This hybrid architecture has direct implications for resource-constrained deployments, where smaller models are preferred due to cost, latency, or deployment constraints.


\section{Conclusion}
\label{section: conclusion}

Our framework, SecureRAG-RTL, utilizes RAG to significantly empower LLMs in hardware vulnerability detection. Our comprehensive evaluation across 18 diverse LLMs demonstrates that SecureRAG-RTL enables substantial improvement in both detection accuracy and vulnerability extraction, particularly for small and medium-scale models. Results show up to an average of 57.14\% increase in detection accuracy and better localization of vulnerabilities in RTL designs. Lightweight models approach near-SoTA performance when guided by effective retrieval, offering a scalable and modular path forward for hardware verification, particularly in resource-constrained environments.


\clearpage
\newpage
\bibliographystyle{ieeetr}
\bibliography{other-bib}

\end{document}